\documentclass[onecolumn, doublespace, 12pt fonts]{aastex6}

\def\d{{\rm d}}
\def\Rg{R_{\rm g}}
\def\cs{c_{\rm s}}
\def\Om{\Omega}
\def\OmK{\Omega_{\rm K}}
\def\vR{v_{R}}

\def\Qvis{Q_{\rm vis}}
\def\Qadv{Q_{\rm adv}}
\def\Qv{Q_{\nu}}
\def\Qz{Q_z}
\def\MBH{M_{\rm BH}}
\def\Msuns{M_{\odot}~{\rm s}^{-1}}
\def\erg{{\rm erg}~{\rm s}^{-1}}

\shorttitle{Vertical advection in hyper-accretion disks}
\shortauthors{Yi et al.}

\begin{document}

\title{Vertical Advection Effects on Hyper-accretion Disks
and Potential Link between Gamma-ray Bursts and Kilonovae}

\author{Tuan Yi\altaffilmark{1}, Wei-Min Gu\altaffilmark{1,3},
Feng Yuan\altaffilmark{2,3}, Tong Liu\altaffilmark{1,3},
Hui-Jun Mu\altaffilmark{1}}

\altaffiltext{1}{Department of Astronomy, Xiamen University, Xiamen,
Fujian 361005, China; guwm@xmu.edu.cn}
\altaffiltext{2}{Shanghai Astronomical Observatory,
Chinese Academy of Sciences, 80 Nandan Road, Shanghai 200030, China}
\altaffiltext{3}{SHAO-XMU Joint Center for Astrophysics,
Xiamen University, Xiamen, Fujian 361005, China}

\begin{abstract}
Recent simulations on super-Eddington accretion flows have shown that,
apart from the diffusion process, the vertical advection based on
magnetic buoyancy can be a more efficient
process to release the trapped photons in the optically thick disk.
As a consequence, the radiative luminosity from the accretion disk
can be far beyond the Eddington value.
Following this spirit, we revisit the structure and radiation
of hyper-accretion disks with mass accretion rates in the range
$10^{-3}\sim 10~\Msuns$.
Our results show that, due to the strong cooling
through the vertical advection, the disk temperature becomes lower
than that in the classic model without the vertical advection process,
and therefore the neutrino luminosity from the disk is lower.
On the other hand, the gamma-ray photons released
through the vertical advection can be extremely super-Eddington.
We argue that the large amount of escaped gamma-ray photons
may have more significant contribution to the primordial fireball
than the neutrino annihilation,
and may hint a link between gamma-ray bursts and kilonovae in the
black hole hyper-accretion scenario.
\end{abstract}


\section{Introduction} \label{sec:intro}

Gamma-ray bursts (GRBs) are the most energetic phenomenon occurred
in the cosmological distance. The luminosity of GRBs can reach
the order of around $10^{50}~ \erg$. There are various models
to explain the energy source of GRBs.
The popular model for the generation of short GRBs is the mergers of
compact binaries \citep[e.g.,][]{Narayan1992,Nakar2007,Berger2014},
either black hole-neutron star binaries or double neutron stars.
And it is commonly believed that the collapse of a massive star
can account for long GRBs \citep[e.g.,][]{Woosley1993,Paczy1998,Woosley2006}.
In both cases, a dense accretion disk is expected to form,
with extremely high temperature and accretion rates.
The neutrino-antineutrino annihilation process is a possible mechanism
to provide energy for the bursts.
The other mechanism based on hyper-accretion disks around black holes
is the well-known Blandford-Znajek (BZ) process \citep[e.g.,][]{Blandford1977,Lee2000}.
Alternatively, a rapidly rotating neutron star with strong magnetic fields
can also be responsible for GRBs \citep[e.g.,][]{Usov1992,Dai1998,Dai2006,Metzger2011,Lv2015}.

In this paper we will focus on the hyper-accreting disks with
extremely high accretion rates ($10^{-3}~\Msuns \la \dot M \la 10~\Msuns$),
where the neutrino radiation may be a dominant cooling mechanism, and therefore
such flows are usually named as neutrino-dominated accretion flows (NDAFs).
\citet{Popham1999} pioneered to study the NDAFs in details,
and proposed that the neutrino annihilation of a hyper-accreting black hole
system can explain GRBs up to $10^{52}~\erg$.
The hyper-accreting black hole system typically has accretion rates around
$0.01 \sim 10~\Msuns$.
The NDAFs have been widely studied on the radial structure and neutrino
radiation
\citep[e.g.,][]{Popham1999,Narayan2001,Di2002,Kohri2002,Kohri2005,
Gu2006,Liu2007,Chen2007,Kawanaka2013,Zalamea2011,Xue2013,Cao2014,Xie2016}, vertical structure and
convection \citep[e.g.,][]{Liu2010,Pan2012,Kawanaka2012,Liu2014,Liu2015a},
and time-dependent variation \citep{Janiuk2004,Janiuk2013}.
The density can reach $10^{8}-10^{12} ~\rm g~cm^{-3}$ in the inner region
of the disk, and the temperature can be up to $10^{10}-10^{11} ~\rm K$.
Such disks can be extremely optically thick,
leading to the trapping of large portion of photons,
which are carried along the radial direction and are eventually absorbed by the
central black hole.
Thus, the neutrino cooling is the dominant cooling process in the inner part.

Recently, whether the neutrino annihilation process can work as the central
engine for GRBs has been investigated in more details
\citep{Liu2015b,Song2015,Song2016}. Such works revealed that
the neutrino annihilation process can account for most GRBs. For some
long-duration GRBs, the central engine is more likely to be the BZ
mechanism rather than the neutrino annihilation since the former is
much more efficient. On the other hand, if the X-ray flares after the prompt
gamma-ray emission are regarded as the reactivity of the central engine,
the neutrino annihilation mechanism may encounter difficulty in interpreting
the X-ray flares. Even including a possible magnetic coupling \citep{Lei2009}
between the inner disk and the central black hole, \citet{Luo2013} showed that
the annihilation mechanism can work for the X-ray flares with duration
$\la 100~{\rm s}$.
However, the annihilation mechanism is unlikely to be
responsible for those long flares with duration $\ga 1000~{\rm s}$,
even though the role of magnetic coupling is included.
More recently, \citet{Mu2016} investigated the central engine of the
extremely late-time X-ray flares with peak time larger than $10^4$ s,
and suggested that neither the neutrino annihilation nor the BZ process
seems to work well. Instead, a fast rotating neutron star with strong bipolar
magnetic fields may account for such flares.

Recent simulations have made significant progress on the
super-Eddington accretion process, including the presence of strong outflows
\citep{Ohsuga2005,Ohsuga2011,Yang2014} and the radiation-powered baryonic jet
\citep{Sadow2015}. In addition, some simulations \citep[e.g.,][]{Sadow2016}
show the anisotropic feature of radiation. More importantly,
the simulation of \citet{Jiang2014} revealed a new energy transport
mechanism in addition to the diffusion, which is named as the vertical
advection. Their simulation results show that, for the super-Eddington
accretion rate $\dot M = 220L_{\rm Edd}/c^2$, the radiative efficiency
is around 4.5\%, which is comparable to the value in a standard thin disk model.
The physical reason is that, a large fraction of photons can escape from
the disk before being advected into the black hole, through the vertical
advection process based on the magnetic buoyancy, which dominates
over the photon diffusion process.

In this work, following the spirit of the vertical advection,
we incorporate the vertical advection process into NDAFs and revisit
the structure and radiation of hyper-accretion disks around
stellar-mass black holes.
The remainder is organized as follows. The basic physics and equations
for our model are described in Section~\ref{sec:equa}.
Numerical results and analyses are presented in Section~\ref{sec:resu}.
Conclusions and discussion are made in Section~\ref{sec:conc}.

\section{Basic Equations} \label{sec:equa}

In this section we describe the basic equations of our model.
We consider a steady state, axisymmetric hyper-accretion disk
around a stellar-mass black hole.
The well-known Paczy{\'n}ski-Wiita potential \citep{PW80}
is adopted, i.e., $\Psi =  - G \MBH / (R-\Rg)$,
where $\MBH$ is the black hole mass
and $\Rg = 2G\MBH / c^2$ is the Schwarzschild radius.
The Keplerian angular velocity can be expressed as
$\OmK =  (G \MBH/R)^{1/2}/(R-\Rg)$.
We use the usual convention of quantities to describe the accretion disk:
the half-thickness of the disk is $H = \cs / \OmK$,
where $\cs = (P/\rho)^{1/2}$ is the isothermal sound speed,
with $P$ being the pressure, and $\rho$ the density.
We adopt the standard Shakura-Sunyaev prescription for
the kinematic viscosity coefficient, i.e. $\nu = \alpha \cs H$.

The basic equations that describe the accretion disk
are the continuity, azimuthal momentum, energy equation,
and the equation of state.
The continuity equation is
\begin{equation}\label{continuity}
\dot{M} = -4 \pi \rho H R \vR \ ,
\end{equation}
where $\vR$ is the radial velocity.
With the Keplerian rotation assumption $\Om = \OmK$,
the azimuthal momentum equation can be simplified as \citep[e.g.,][]{Gu2006,Liu2007}:
\begin{equation}\label{momentum}
\vR = -\alpha \cs \frac{H}{R} f^{-1} g \ ,
\end{equation}
where $f = 1 - j / \OmK  R^2$, $g = - \d \ln \OmK / \d \ln R$,
and $j$ represents the specific angular momentum per unit mass
accreted by the black hole.
The equation of state takes the form \citep[e.g.,][]{Di2002}
\begin{equation}\label{pressure}
P = \frac{\rho k_{{\rm B}} T}{m_{\rm p}} \left( \frac{1+3X_{\rm nuc}}{4} \right)
+ \frac{11}{12}aT^4
+ \frac{2 \pi h c}{3} (\frac{3}{8 \pi m_{\rm p}}
\frac{\rho}{\mu_{e}})^{\frac{4}{3}} + \frac{u_{\nu}}{3} \ ,
\end{equation}
where the four terms on the right-hand side are the gas pressure,
radiation pressure of photons, degeneracy pressure of electrons,
and radiation pressure of neutrinos, respectively.

The energy equation is written as
\begin{equation}\label{energy}
\Qvis = \Qadv  + \Qz + \Qv \ .
\end{equation}
The above equation shows the balance between the viscous heating
and the cooling by radial advection, vertical advection, and neutrino
radiation. The viscous heating rate $\Qvis$ and the advective cooling
rate $\Qadv$ for a half-disk above or below the equator are expressed as
\citep[e.g.,][]{Di2002,Gu2006}
\begin{equation}
\Qvis  = \frac{1}{4\pi} \dot M \Omega^2 f g \ ,
\end{equation}
\begin{equation}\label{Qadv}
\Qadv = - \xi \vR \frac{H}{R}T
( \frac{11}{3}aT^3 + \frac{3}{2} \frac{\rho k_{{\rm B}}}{m_{\rm p}}
\frac{1+3X_{\rm nuc}}{4} + \frac{4}{3}\frac{u_{\nu}}{T} ) \ ,
\end{equation}
where $T$ is the temperature, $s$ is the specific entropy,
$u_{\nu}$ is the neutrino energy density, and $\xi$ is taken to be $1$.
$X_{\rm nuc}$ is the mass fraction of free nucleons \citep[e.g.,][]{Kohri2005}:
\begin{equation}
X_{\rm nuc} = {\rm Min} [1,~20.13 \rho_{10}^{-3/4} T_{11}^{9/8}
\exp (-0.61/T_{11})] \ ,
\end{equation}
where $\rho_{10}= \rho /10^{10} ~\rm g~cm^{-3}$ and
$T_{11} = T / 10^{11} ~\rm K$.
The quantity $ \Qv $ is the cooling rate due to the neutrino radiation.
We adopt a bridging formula for calculating $\Qv$ as shown in \citet{Di2002} and \citet{Liu2007}.

The main difference from previous works is that
the vertical advection term $\Qz$ is taken into account in our work,
which is written as
\begin{equation}
\Qz = \overline{V_z} ( u_{\rm ph} + u_{\nu} + u_{\rm gas} ) \ ,
\end{equation}
where $u_{\rm ph}$ is the energy density of photons,
$u_{\nu}$ is the energy density of neutrinos,
$u_{\rm gas}$ is the energy density of the gas.
In our calculation the third term $\overline{V_z} u_{\rm gas}$
is dropped since the escaped gas through
the magnetic buoyancy can be negligible.
The quantity $\overline{V_z}$ is the averaged velocity of the vertical
advection process, which can be simply written as
\begin{equation}
\overline{V_z} = \lambda \cs \ ,
\end{equation}
where $\lambda$ is a dimensionless parameter.
By comparing the typical vertical velocity $\overline{V_z}$
in the simulation results
\citep[Figure 14 of][]{Jiang2014} and the theoretical estimate of sound speed,
we take $\lambda = 0.1$ for our numerical calculations.
The vertical advection term describes the released photons and neutrinos
due to the magnetic buoyancy, which can dominate over
normal diffusion process.

Equations~(\ref{continuity})-(\ref{energy}) can be solved if the parameters
$\MBH$, $\dot{M}$, $\alpha$, and $j$ are given.
We consider a stellar-mass black hole with $\MBH = 3 M_{\odot}$.
The viscous parameter $\alpha = 0.02$ is taken from the simulation
results \citep{Hirose2009}.
The specific angular momentum $j = 1.83 c\Rg$ is just a little less than
the Keplerian angular momentum at the marginally stable orbit, i.e.,
$l_{\rm K}|_{3 \Rg} = 1.837c\Rg$.
Our study focuses on the solutions in the range $3\sim 10^3~\Rg$.

\section{Numerical Results} \label{sec:resu}

\begin{figure}
\figurenum{1}
\plotone{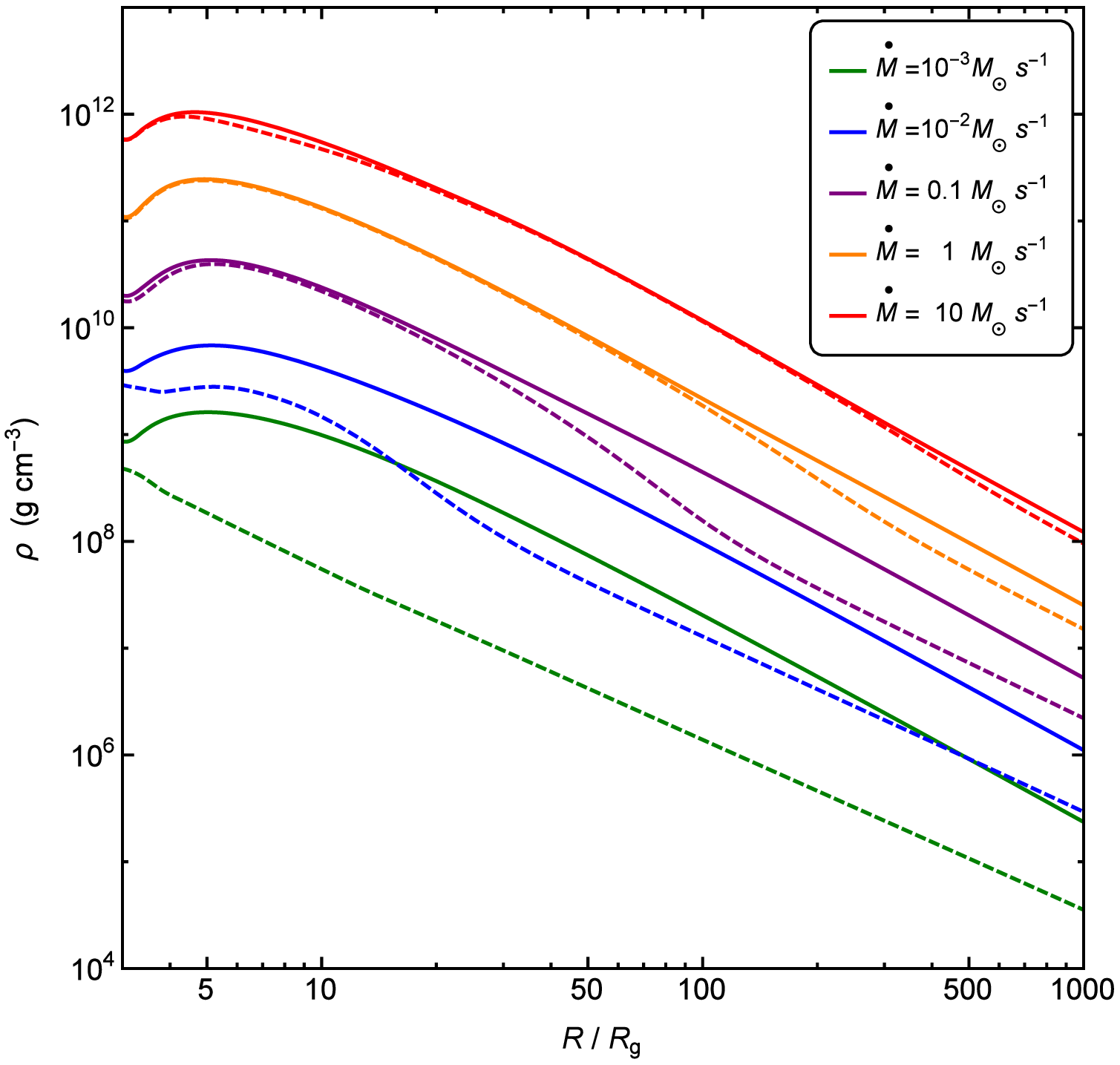}
\caption{
Radial profiles of density for five typical mass accretion rates,
where the solid lines represent the results with vertical advection process
and the dashed lines represent the results without such process.
}
\label{F:01}
\end{figure}

\begin{figure}
\figurenum{2}
\plotone{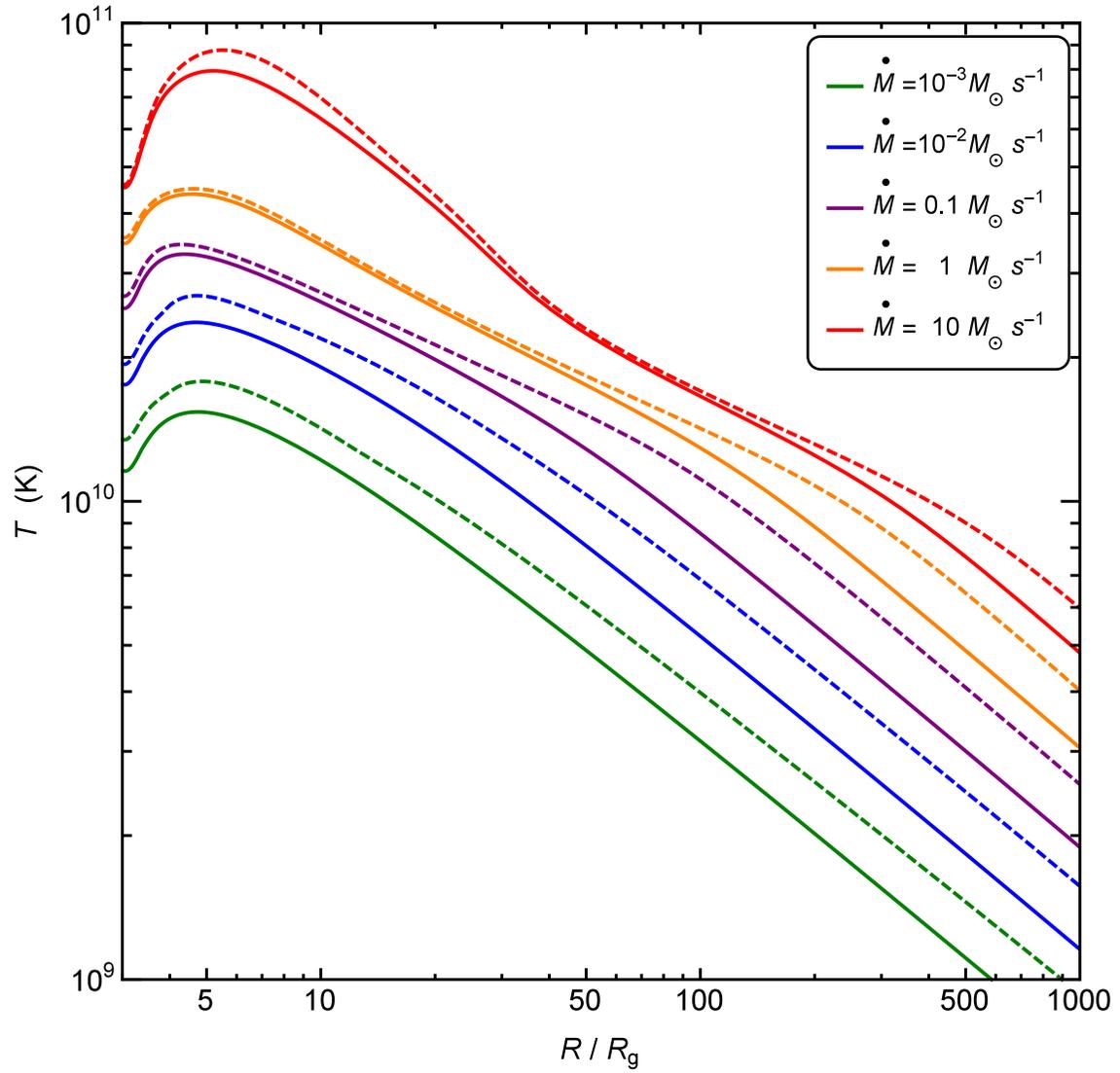}
\caption{
Same as Figure~1 but for the temperature profiles.
}
\label{F:02}
\end{figure}

In this section we present our numerical results and the analyses of
the physics behind these results.
The calculation reveals that the vertical advection process has essential
effects on the structure and radiation of the disk.
First, the radial profiles of mass density and temperature are investigated
for two cases, i.e., with and without the vertical advection process.
The radial profiles of density $\rho$ are shown in Figure~\ref{F:01},
where the solid (dashed) lines correspond to the results with (without)
the vertical advection process. The five typical mass accretion rates
are $\dot M = 10^{-3}$, $10^{-2}$, $0.1$, $1$, and $10~\Msuns$, which are
shown by different colors.
It is seen from Figure~\ref{F:01} that, for a certain radius $R$ and a given
$\dot M$, the density $\rho$ in the disk with vertical advection process
is significantly higher than that without such a process, particularly for
low accretion rates such as $\dot M = 10^{-3} ~\Msuns$.
The radial profiles of temperature $T$ are shown in Figure~\ref{F:02}, where
the explanation of different color and types of lines is the same as in
Figure~\ref{F:01}.
It is seen that the temperature of the disk with the vertical advection
is generally lower than that without the advection.

The physical understanding of the above difference in density and temperature
is as follows. Since a large fraction of the trapped photons can be released
through the vertical advection process, the radiative cooling
through such a process is efficient. As a consequence, the temperature $T$
together with the total pressure $P$, the vertical height $H/R$, and the radial
velocity $v_R$ will decrease, whereas the mass density $\rho$ will increase.
For relatively low accretion rates such as $\dot M = 10^{-3}~ \Msuns$,
the radiative cooling of neutrinos is quite inefficient, so the
effects of vertical advection may be more significant, as indicated by
Equation~(\ref{energy}).

\begin{figure}
\figurenum{3}
\plotone{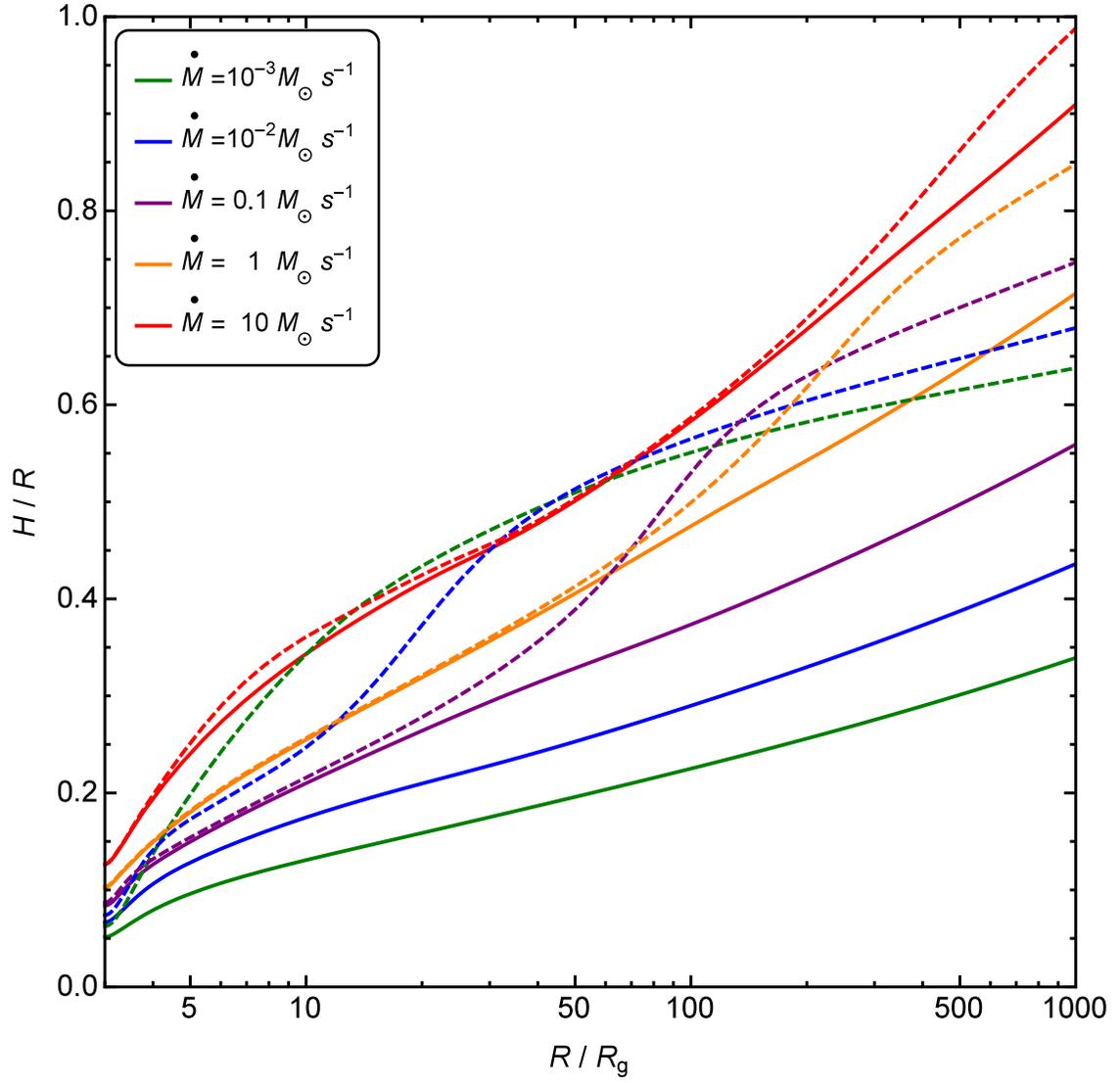}
\caption{
Same as Figure~1 but for the vertical height profiles.
}
\label{F:03}
\end{figure}

Figure~\ref{F:03} shows the radial profiles of vertical scale height of the disk.
It is seen that the relative height $H/R$ of the disk with vertical advection
is significantly thinner than that without the vertical advection,
particularly for low accretion rates.
The physical reason is mentioned above, which is
related to the decrease of temperature and pressure.
The decrease of vertical height also implies the decrease of sound speed
and therefore the decrease of radial velocity, as inferred by
Equation~(\ref{momentum}).

\begin{figure}
\figurenum{4}
\plotone{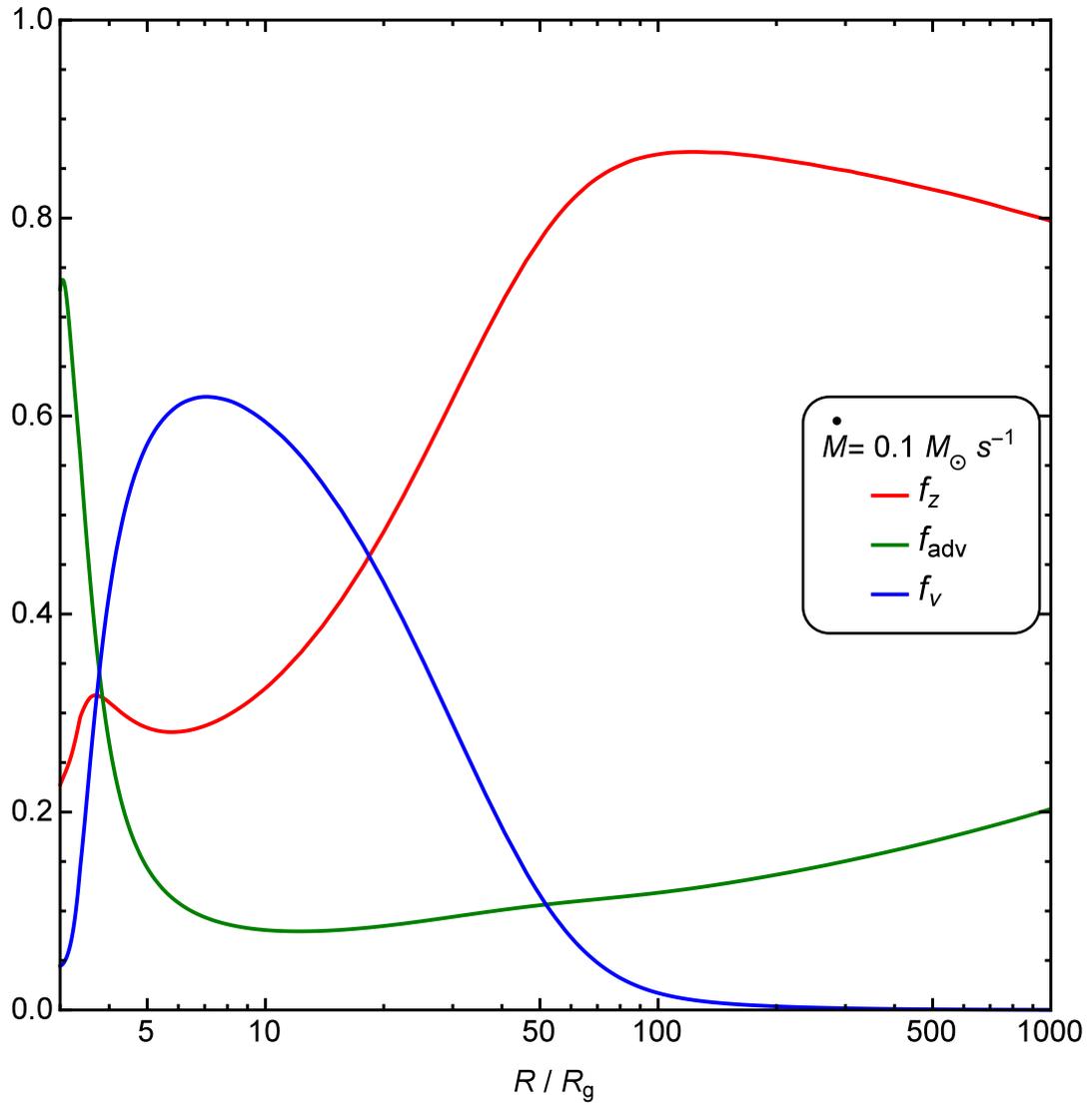}
\caption{
Radial profiles of energy components for a typical mass accretion rate
$\dot M = 0.1~\Msuns$.
}
\label{F:04}
\end{figure}

For a typical mass accretion rate $\dot M = 0.1~\Msuns$,
Figure~\ref{F:04} shows the radial profiles of energy components,
where the three dimensionless factors are
$f_z = Q_z/Q_{\rm vis}$ (red line),
$f_{\rm adv} = Q_{\rm adv}/Q_{\rm vis}$ (green line),
and $f_{\nu} = Q_{\nu}/Q_{\rm vis}$ (blue line).
It is seen that in the most inner region ($R \la 4\Rg$) the radial advection
is the dominant cooling mechanism.
For the outer part of the disk ($R \ga 20\Rg$) the energy transport
through the vertical advection becomes dominant.
For the region $4\Rg \la R \la 20\Rg$ the neutrino cooling may dominate over
the other two mechanisms.
The red line implies that the photon radiation through the vertical
advection process can reach a large fraction of the total released
gravitational energy, and therefore the photon luminosity can be extremely
super-Eddington.
We will investigate the corresponding photon luminosity in Figure~\ref{F:06}.

\begin{figure}
\figurenum{5}
\plotone{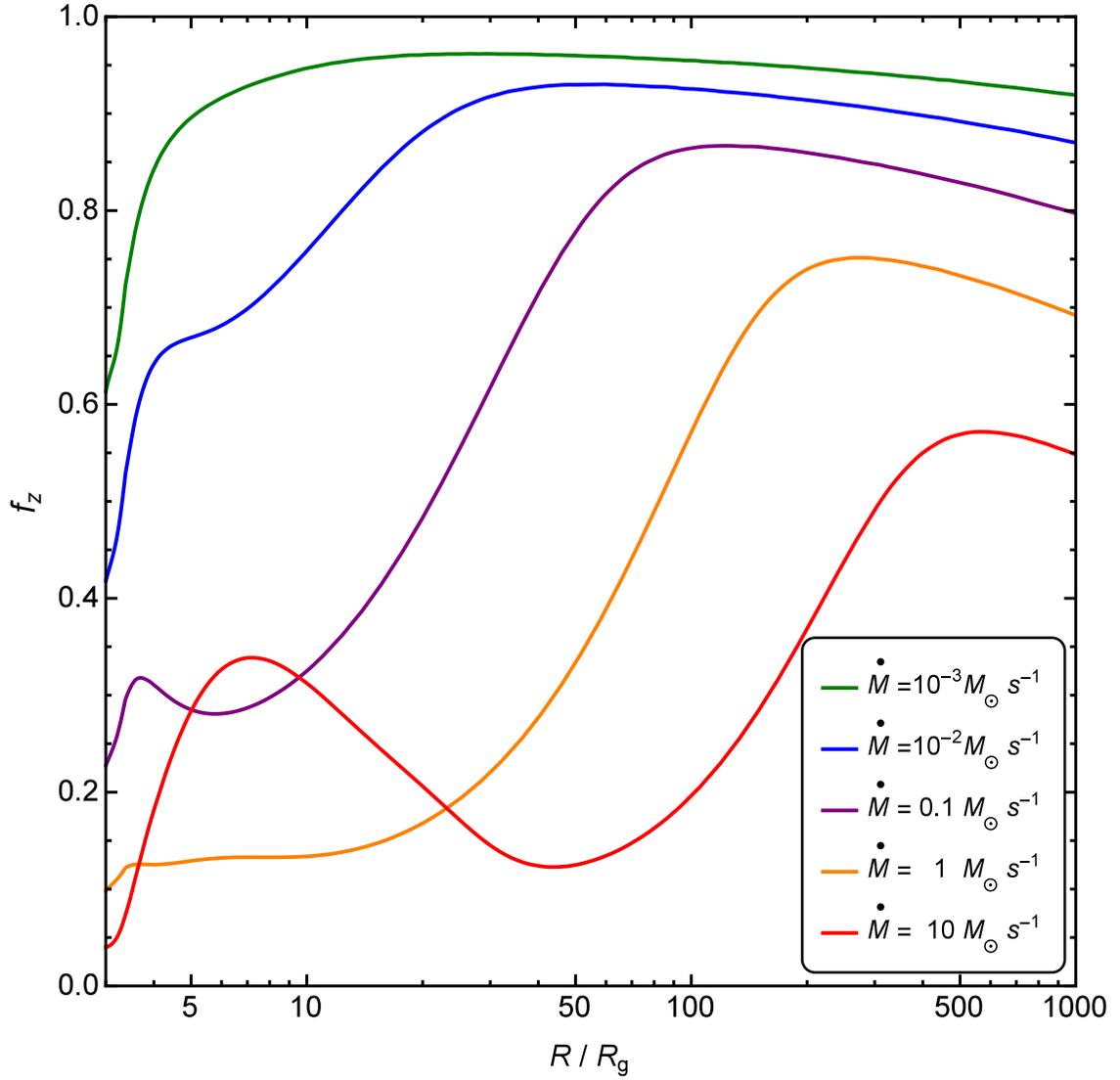}
\caption{
Radial profiles of the ratio of the vertical energy advection rate $Q_z$
to the viscous heating rate $Q_{\rm vis}$ for five typical mass accretion rates.
}
\label{F:05}
\end{figure}

\begin{figure}
\figurenum{6}
\plotone{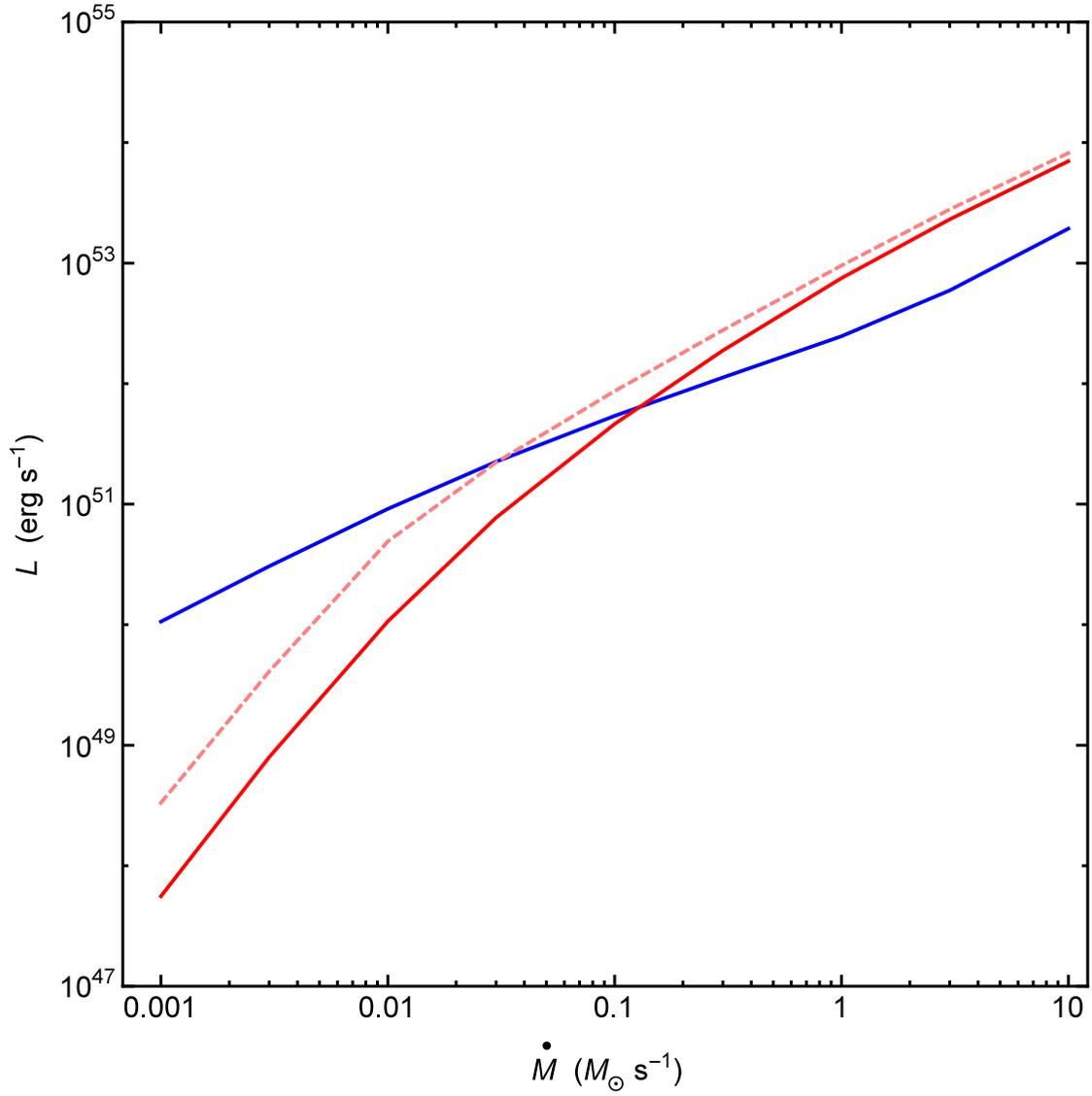}
\caption{
Variation of the luminosity with the mass accretion rate, where the red
solid (dashed) line corresponds to the neutrino luminosity with (without)
the vertical advection process, and the blue line corresponds to the
gamma-ray luminosity due to the released photons through the vertical
advection process.
}
\label{F:06}
\end{figure}

Our main focus is the energy transport through vertical advection process.
Figure~\ref{F:05} shows the radial profiles of the dimensionless cooling rate
due to the vertical advection, i.e., $f_z = Q_z/Q_{\rm vis}$,
where five typical accretion rates are adopted.
It is seen that $f_z$ generally decreases with
increasing $\dot M$. The physical reason is that the neutrino cooling
is less important for relative low accretion rates.
For the highest accretion rate with $\dot M = 10~\Msuns$, the red line shows that
there exists a big bump in the inner region ($\la 20\Rg$).
The physics for this bump is
that the neutrino cooling is again less significant since the inner disk
is optically thick to the neutrinos.

Finally, we calculate the neutrino luminosity and the photon luminosity
of the accretion disk.
The neutrino luminosity is derived by the integration of the whole disk:
\begin{equation}
L_{\nu} = \int^{R_{\rm out}}_{R_{\rm in}}
4 \pi R \cdot (Q_{\nu} + \overline{V_z} u_{\nu}) \ \d R \ ,
\end{equation}
where $L_{\nu}$ includes the contributions from the direct neutrino radiation
$Q_{\nu}$ and the vertical advection process on neutrinos
$\overline{V_z} u_{\nu}$.
Actually, the latter is negligible except for extremely high accretion rates
$\dot M \gg 1~\Msuns$.
The variation of neutrino luminosity with mass accretion rates is shown in
Figure~\ref{F:06}, where the red solid line corresponds to the neutrino
luminosity with the vertical advection process, whereas the red dashed line
corresponds to the neutrino luminosity without the process. It is seen
that the red solid line is under the red dashed line, which means that
the neutrino luminosity with the vertical advection is lower. The physical
reason is as follows. The neutrino radiation is more sensitive to
the temperature than the density. As shown by Figure~\ref{F:02},
the disk temperature is lower for the case with the vertical advection.
Thus, the corresponding $Q_{\nu}$ and $L_{\nu}$ in our cases are lower
than those without the vertical advection.

The variation of photon luminosity $L_{\rm ph}$ is shown by the blue solid
line in Figure~\ref{F:06}, where $L_{\rm ph}$ is calculated by
\begin{equation}
L_{\rm ph} = \int^{R_{\rm out}}_{R_{\rm in}}
4 \pi R \cdot \overline{V_z} u_{\rm ph} \ \d R \ .
\end{equation}
It is seen that $L_{\rm ph}$ is in the range $10^{50}\sim 10^{53}~\erg$ for
$0.001 \leqslant \dot m \leqslant 10$, which is more than ten orders of
magnitude higher than the Eddington luminosity. The released photons are mainly
in the gamma-ray band according to the thermal radiation of the inner disk
with $10^{10}{\rm K} < T < 10^{11}{\rm K}$, as shown by Figure~\ref{F:02}.
The huge amount of gamma-ray photons escape from the optically thick disk
through the vertical advection process, which is much more efficient than
the diffusion process. Such an extremely high photon radiation should have
observational effects. We will have a discussion on that in next section.

\section{Conclusions and discussion} \label{sec:conc}

In this work, we have studied the structure and radiation of hyper-accretion
flows around stellar-mass black holes by taking into the role of
vertical advection process. By the comparison of our results with the
classic NDAF solutions, we have shown that
the density is higher, the temperature is lower, and the vertical
height is thinner in our solutions.
The physical reason is that a large fraction of photons can escape
from the optically thick disk through the vertical advection process.
As a consequence, the neutrino luminosity from the disk is
decreased. Thus, even without calculating the neutrino annihilation
luminosity, we can conclude that the annihilation mechanism cannot be
responsible for the long-duration GRBs and X-ray flares.

We would point out that outflows are not taken into consideration in the
present work. However,
outflows are believed to generally exist in accretion flows.
Recent MHD simulations have shown that outflows
exist both in optically thin flows \citep[][]{Yuan2012a,Yuan2012b}
and optically thick flows \citep[][]{Jiang2014,Sadow2015,Sadow2016}.
From the observational view,
\citet{Wang2013} reveals that more than 99\% of the accreted mass escape
from the accretion flow by outflows in our Galactic center.
Based on the energy balance argument, \citet{Gu2015} shows that
the outflow is inevitable for the accretion flows where the radiative cooling
is far below the viscous heating, no matter the flow is optically thin
or thick.
Thus, outflows may work as another process to help the trapped photons
to escape \citep{Shen2015}, and will also have effects on the structure
and neutrino radiation of the accretion flow.
Such a mechanism is not included in the current work.

Our calculations are based on the relation
$\overline{V_z} = \lambda \cs$ with $\lambda = 0.1$.
As mentioned in Section~2, the value for $\lambda$ is adopted following
the simulation results for $\dot M = 220 L_{\rm Edd}/c^2$ \citep{Jiang2014}.
In our case, the mass accretion rate is higher for more than ten
orders of magnitude.
Then, a key question may exist whether the vertical advection due to
the magnetic buoyancy can also work for such hyper-accretion systems.
In our opinion, the radiation pressure is always dominant up to
$\dot M \la 0.1~\Msuns$ or for the outer part of even higher accretion rates.
Thus, such a mechanism seems to be an efficient process.
On the other hand, even for the case that the parameter $\lambda$ is
significantly smaller than $0.1$ in the hyper-accretion case, such as
several orders of magnitude smaller, the released gamma-ray photons may
still be extremely super-Eddington and the potential application is significant.

In this work we have assumed $\alpha = 0.02$ according to the simulation
results of \citet{Hirose2009}. However, other simulations may provide
different values for $\alpha$. As shown by \citet{Yuan2014}, such a value
may be related to the magnitude of net magnetic flux in the simulations.
The values of $\alpha$ may also have significant effects on the energy transport
of the vertical advection. As Equations~(\ref{momentum}) and (\ref{Qadv}) imply,
$\vR$ is proportional to $\alpha$ and $\Qadv$ is proportional to
$\vR$ and therefore $\alpha$. Thus, we can expect that, for a larger
value of $\alpha$, the advective cooling rate $\Qadv$ can significantly
increase and therefore the cooling rate due to the vertical advection
$Q_z$ will decrease according to the energy balance of Equation~(\ref{energy}).
Nevertheless, the luminosity related to the radial integration of $Q_z$
will still be hyper-Eddington even though $Q_z$ may be lower than
$\Qadv$ for a large range of radii.

\begin{figure}
\figurenum{7}
\plotone{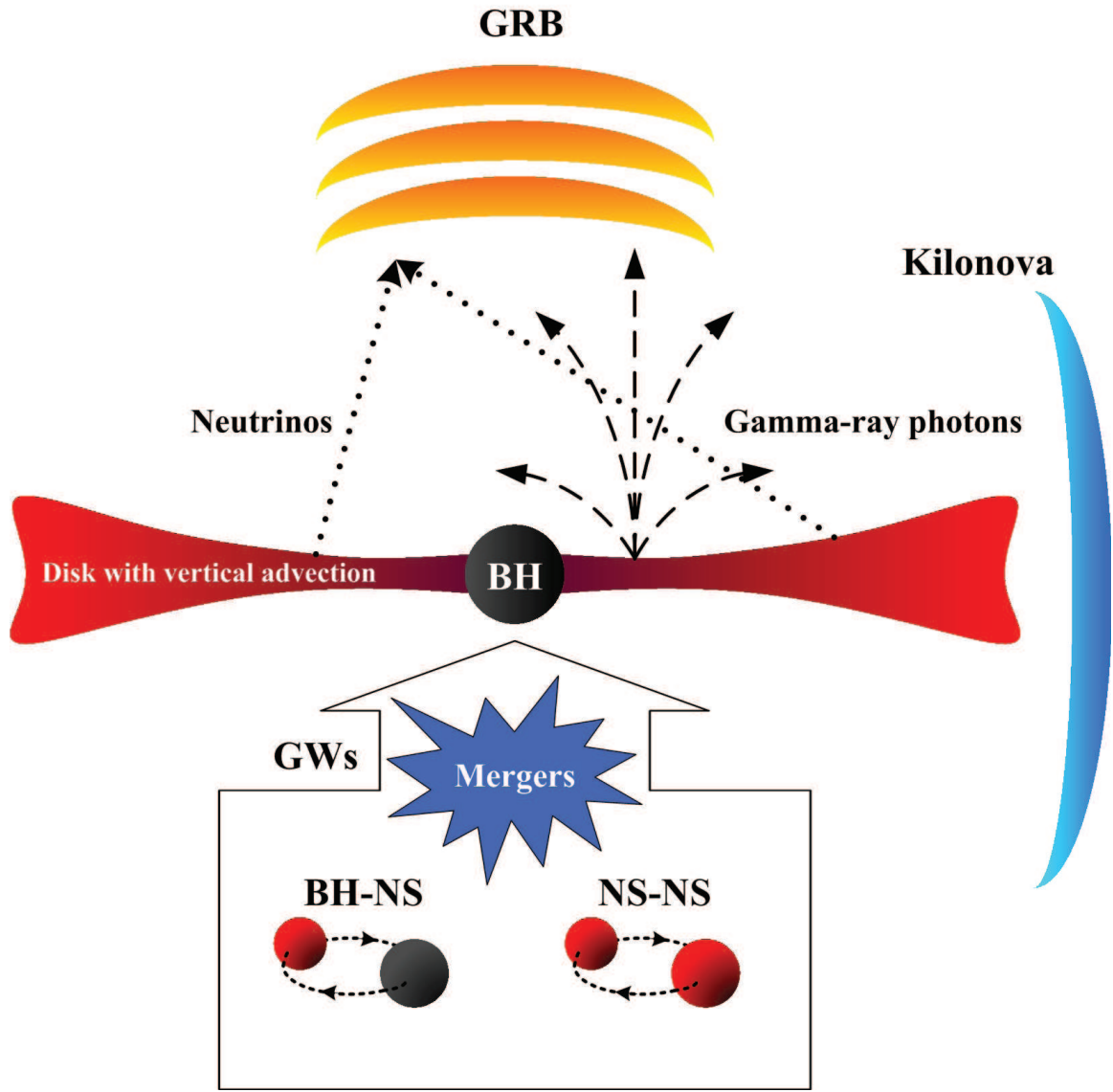}
\caption{
Illustration of the association of short GRBs, kilonovae,
and gravitational wave events.}
\label{F:07}
\end{figure}

Our results of extremely super-Eddington luminosity of gamma-ray emission
can also be generally applied to short GRBs.
It is commonly believed that short GRBs originate from the merger of
compact objects, i.e., the black hole-neutron star binary or
the binary with double neutron stars.
More importantly, the merger is a significant source of
gravitational wave event.
Thus, the released high-energy photons may have significant contribution
to an electromagnetic counterpart for the gravitational wave event,
such as kilonovae.
Kilonovae have been widely studied in recent years
\citep[e.g.,][]{Li98,Metzger2012,Jin2013,Yu2013,Gao2015,Kasen2015,
Kawaguchi2016,Metzger2016,Fernandez2016a}.
In our case, a new picture is shown by Figure~\ref{F:07}.
It is seen from this figure that, either the merger of a black hole
and a neutron star or the merger of two neutron stars may result in
a gravitational wave event and a black hole hyper-accretion disk.
According to our study in this work,
most gamma-ray photons escaping from the direction perpendicular to
the equatorial plane together with the neutrino annihilation
contribute to the thermal fireball,
and the remanent escaped photons diverge from other directions to trigger
non-thermal emission due to diffusion into ambient environment.
Obviously, such a progenitor of kilonovae is quite different from
the origin from disk wind \citep[e.g.,][]{Metzger2012,Kasen2015}
or magnetar wind \citep[e.g.,][]{Yu2013,Gao2015}. 
In addition, a faint gamma-ray thermal component may
exist owing to the large amount of escaped thermal gamma-ray photons
from the disk, which may have potential contribution to the thermal component
of the prompt gamma-ray emission \citep{Abdo2009,Zhang2016}.
Moreover, the neutrino annihilation mechanism or the BZ mechanism based on
the accumulation of magnetic fields through the hyper-accretion process
will work as the main central engine for the GRB.
In summary, in our scenario of black hole hyper-accretion with
vertical advection process, short GRBs, kilonovae, and gravitational
wave events can be naturally blended together \citep[e.g.,][]{Fernandez2016b}.

\acknowledgments

This work was supported by the National Basic Research Program of China
(973 Program) under grants 2014CB845800,
the National Natural Science Foundation of China under grants 11573023,
11473022, 11573051, 11633006, and 11333004,
the National Program on Key Research and Development Project of China
(Grant No. 2016YFA0400704),
the Key Research Program of Frontier Sciences of CAS (No. QYZDJ-SSW-SYS008),
and the CAS Open Research Program of Key Laboratory for the Structure and
Evolution of Celestial Objects under grant OP201503.

\clearpage

\clearpage

\clearpage

\clearpage

\clearpage

\clearpage

\clearpage

\end{document}